\pdfoutput=1
\documentclass[aps,prd,nofootinbib,twocolumn,preprintnumbers,balancelastpage]{revtex4}
\usepackage{amssymb,amsmath,latexsym,graphics, graphicx,epsfig,multirow,comment,hyperref,appendix, verbatim} 
\usepackage{pifont}
\usepackage{lipsum}

\usepackage{slashed}
\usepackage{subfigure}
\usepackage{feynmp}
\usepackage{graphicx}
\usepackage{amsfonts}
\usepackage{color}
\newcommand{\be}{\begin{equation}}
\newcommand{\ee}{\end{equation}}
\newcommand{\bea}{\begin{eqnarray}}
\newcommand{\eea}{\end{eqnarray}}

\newcommand{\ba}{\begin{eqnarray}}
\newcommand{\ea}{\end{eqnarray}}

\begin{document}
\title[]{Constraints on Dark Matter Protohalos in \\Effective Theories and Neutrinophilic Dark Matter
}

\author{Ian M. Shoemaker}
\affiliation{Theoretical Division T-2, MS B285, Los
Alamos
  National Laboratory, Los Alamos, NM 87545, USA}

\begin{abstract}

The mass of primordial dark matter (DM) protohalos remains unknown. However, the missing satellites problem may be an indication that they are quite large. In this paper, we use effective field theory to map constraints on dark matter-SM interactions into limits on the mass of DM protohalos.  Given that leptons remain in the thermal bath until late times, we focus on their interactions with DM. To illustrate the method, we use the null results of LEP missing energy searches along with Fermi-LAT searches for DM annihilation in nearby dwarf galaxies, to derive limits on the protohalo mass, $\lesssim (10^{-6}-10^{-1})~M_{\odot}$, with the range depending on the DM mass and the operator.  Thus, if DM is to remain thermally coupled until late times and account for the missing satellites, charged lepton interactions are insufficient.  This motivates neutrinophilic DM, which can have protohalo masses orders of magnitude larger, with constraints arising from Planck, IceCube and unpublished Super-K data.  We show that effective neutrinophilic models offer a viable solution to the missing satellites problem for sub-GeV DM masses with larger than WIMP-sized annihilation cross sections.

\end{abstract}

\maketitle

\preprint{LA-UR-13-22524}

\newpage

\section{Introduction}

One of the simplest particle candidates for DM is the thermally produced weakly-interacting massive particle (WIMP)~\cite{Jungman:1995df}.   WIMPs with sufficient interaction strength are kept in chemical (number-changing) and kinetic (momentum-changing) equilibrium with the thermal bath. Thanks to the Universe's expansion however, this state of affairs is lost once the rate of WIMP-bath scattering is comparable to the rate of Hubble expansion. Number-changing processes ($XX \leftrightarrow ff$) are the first to go out of equilibrium, fixing the comoving WIMP number density at a value determined by the annihilation cross section. For a wide range of masses, the observed DM density is obtained with a thermally averaged cross section of $\langle \sigma v \rangle \simeq 3\times 10^{-26}~{\rm cm}^{3}{\rm s}^{-1}$ ($6\times 10^{-26}~{\rm cm}^{3}{\rm s}^{-1}$) for Majorana (Dirac) particles. If, like baryonic matter, DM possess a nonzero particle-antiparticle asymmetry, an analogous story plays out and the correct relic abundance is attained for cross sections greater than the WIMP value~\cite{Graesser:2011wi}.

Typically the WIMP story ends with chemical decoupling. However, as long as elastic, momentum-changing scattering processes ($X f \leftrightarrow Xf$) continue, the WIMP population will remain in approximate thermal equilibrium. Eventually of course, even elastic scattering fails to keep up with Hubble expansion and {\it kinetic} decoupling of the WIMP population occurs. This epoch imprints a cutoff, $M_{cut}$, in the power spectrum, which determines the size of the first and smallest DM structures~\cite{Green:2005fa,Loeb:2005pm,Bertschinger:2006nq}. Moreover, it is the interactions that remain in equilibrium latest that determine this cutoff scale.

These first and smallest gravitationally bound DM structures remain of great interest.  Should they survive until present, they can impact searches of DM in a number of ways (for a nice review, see e.g. \cite{Koushiappas:2009du}). First, if DM annihilates at large rates to photons, these protohalos are sufficiently dense to be visible in the gamma-ray sky. This has already prompted the Fermi collaboration to search for unassociated sources of gamma-rays, so far with only null results~\cite{Ackermann:2012nb}. In the case of DM direct detection experiments, the rate is on average markedly diminished compared to a smooth halo since substantial mass in substructure reduces the DM volume fraction of the Galaxy~\cite{Kamionkowski:2008vw}.  Moreover, although it is estimated to be unlikely, were the Earth to pass through a dense protohalo, a dramatic increase in the DM flux could be probed by direct detection experiments~\cite{Koushiappas:2009du}.  Similarly, one could use a combination of direct detection data and the $\nu$ signal from DM annihilation in the Sun as a probe of the local DM substructure~\cite{Koushiappas:2009ee}. Lastly, it may be possible to detect DM protohalos directly through their gravitational effects. For example, it has recently it has been pointed out that DM substructure can produce a frequency shift in pulsar timing measurements that can be probed at the Square Kilometer Array~\cite{Baghram:2011is}. Moreover, both strong gravitational lensing~\cite{Moustakas:2009na} and ``nanolensing''~\cite{Chen:2010ae} may with future data yield an orthogonal experimental handle on DM subhalos.

Intriguingly however, we may at present already be faced with observations that indicate that these protohalos are quite massive.  In particular, there are many fewer dwarf galaxy-size structures orbiting the Milky Way than expected from N-body simulations of cold dark matter structure formation~\cite{Moore:1999nt,Klypin:1999uc}. This so-called ``missing satellite problem'' could be accounted for from: (1) baryonic physics that may both flatten the DM density of satellite galaxies and introduce strong tidal stripping that would decrease subhalo survivability~\cite{Penarrubia:2010jk,Zolotov:2012xd,Brooks:2012ah} or (2) a much larger cutoff in the power spectrum of DM than is conventionally assumed. If the second option is realized in Nature, it may imply either that DM is warm (e.g. a keV sterile neutrino~\cite{Kusenko:2009up}) with a large free-streaming length, or that it remained in kinetic equilibrium until late times~\cite{Aarssen:2012fx}. 

If late kinetic decoupling is the answer, DM must have fairly strong interactions with the thermal bath. These same interactions can be probed by terrestrial experiments and astrophysical observations.  In this paper we have two related aims: (1) to estimate how the constraints on DM's coupling to SM particles translate into limits on the protohalo mass; and (2) to examine what model ingredients are needed to accommodate the large protohalo masses the missing satellite problem may hint at.

However, not all particles in the thermal bath contribute equally to DM's kinetic decoupling.   Given that the number density of hadrons becomes small after the QCD phase transition, the coupling of DM to leptons can be much more important in setting the protohalo scale. We shall henceforth focus on the coupling of DM to leptons, and set out to determine what the current experimental sensitivity to these interactions can say about the allowed size of DM protohalos. We will see that leptophilic and neutrinophilic models of DM yield very different results than the MSSM neutralino~\cite{Schmid:1998mx,Chen:2001jz,Bringmann:2006mu,Profumo:2006bv,Bringmann:2009vf,Gondolo:2012vh,Cornell:2012tb}, or DM models with quark-only couplings~\cite{Gondolo:2012vh}. 

To remain as model-independent as possible we assume that lepton-DM interactions can be described using effective field theory.  This is an excellent approximation at the low temperatures relevant for kinetic decoupling, though it of course becomes circumspect at high-energy colliders.  We will examine a handful of qualitatively distinct operators. In all cases, the derived limits imply that DM-lepton scattering cannot be a solution to the missing satellites problem. This motivates neutrinophilic dark matter ($\nu$DM), where the DM-neutrino interaction is much stronger than the DM-charged lepton interaction.  With Planck, IceCube, Super-K being the only relevant experimental probes, we find that for $\nu$DM to accommodate the missing satellite problem thermal DM must be sub-GeV in mass and with a larger than WIMP-sized annihilation cross section. Such a scenario can easily obtain the correct thermal relic abundance if dark matter carries a nonzero asymmetry~\cite{Graesser:2011wi}. Thus future CMB and neutrino telescope data will be critical in determining whether or not $\nu$DM is the solution to the missing satellites problem.

The remainder of this paper is organized as follows. In Sec.~\ref{types} we discuss our model set-up and derive the relevant annihilation and scattering cross sections. In Sec.~\ref{exp} we briefly review the experimental searches relevant for our study. Next, in Sec.~\ref{KD} we review an analytic framework for the kinetic decoupling and free-streaming processes.  In Sec.~\ref{lepDM} we present the main results of this paper in the form of approximate exclusion plots in the protohalo-DM mass plane. In Sec.~\ref{conc} we conclude with a discussion of possible extensions and future probes of $\nu$DM.

\section{Effective Dark Matter-Lepton Interactions}
Given that the charged leptons and neutrinos are among the last SM particles remaining in the thermal bath, we will be interested in the coupling of DM to these particles. LHC and direct detection data can be used to constrain the epoch of DM's kinetic decoupling from quarks. This has been studied recently with the authors of~\cite{Gondolo:2012vh} finding $M_{cut} \lesssim 10^{-6}~M_{\odot}$, with a weak dependence on the Lorentz structure of the operator.   Thus, quark-DM interactions cannot offer a solution to the missing satellites problem.  Moreover, it is important to stress {\it a priori} that leptons and DM to can continue to exchange momentum long after the quark number density  is strongly Boltzmann-suppressed after the QCD phase transition.

\begin{figure*}[t] 
\begin{center}
\includegraphics[width=0.95\columnwidth]{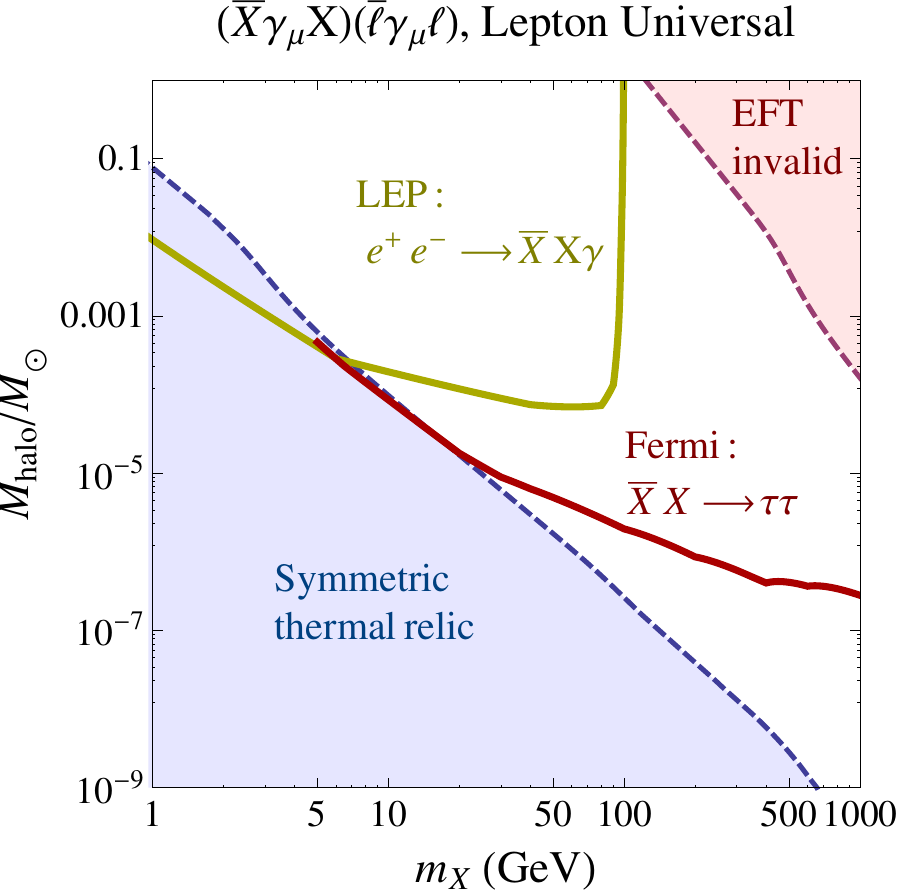}\\
\includegraphics[width=0.95\columnwidth]{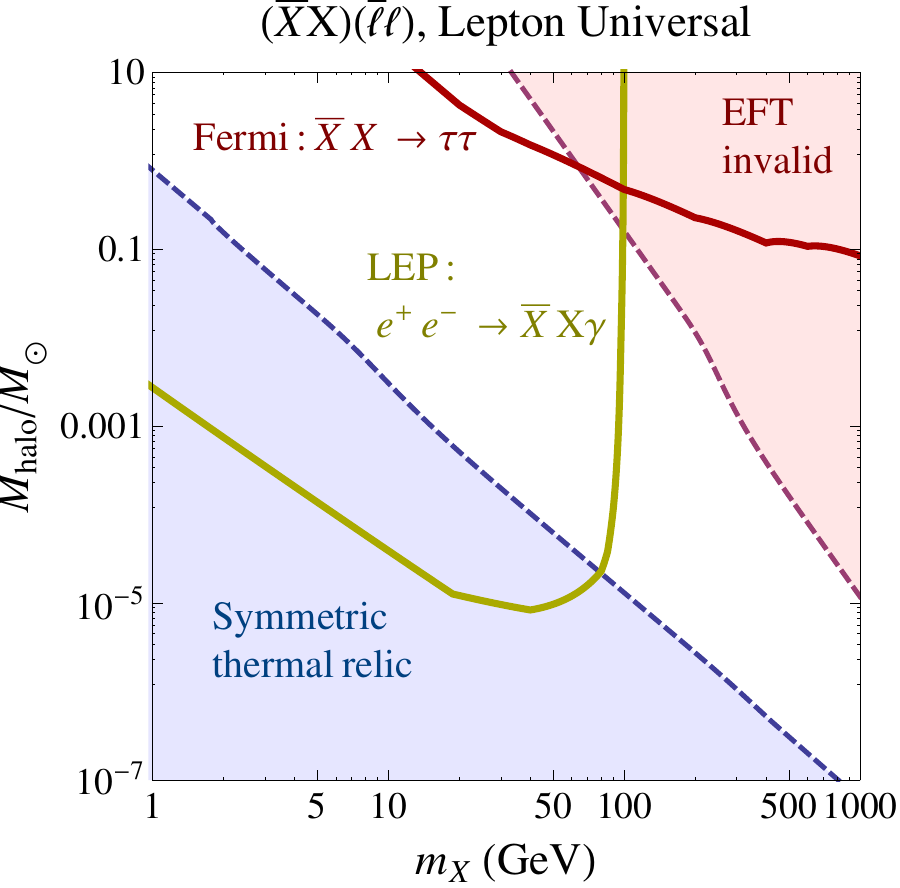}
 \includegraphics[width=0.95\columnwidth]{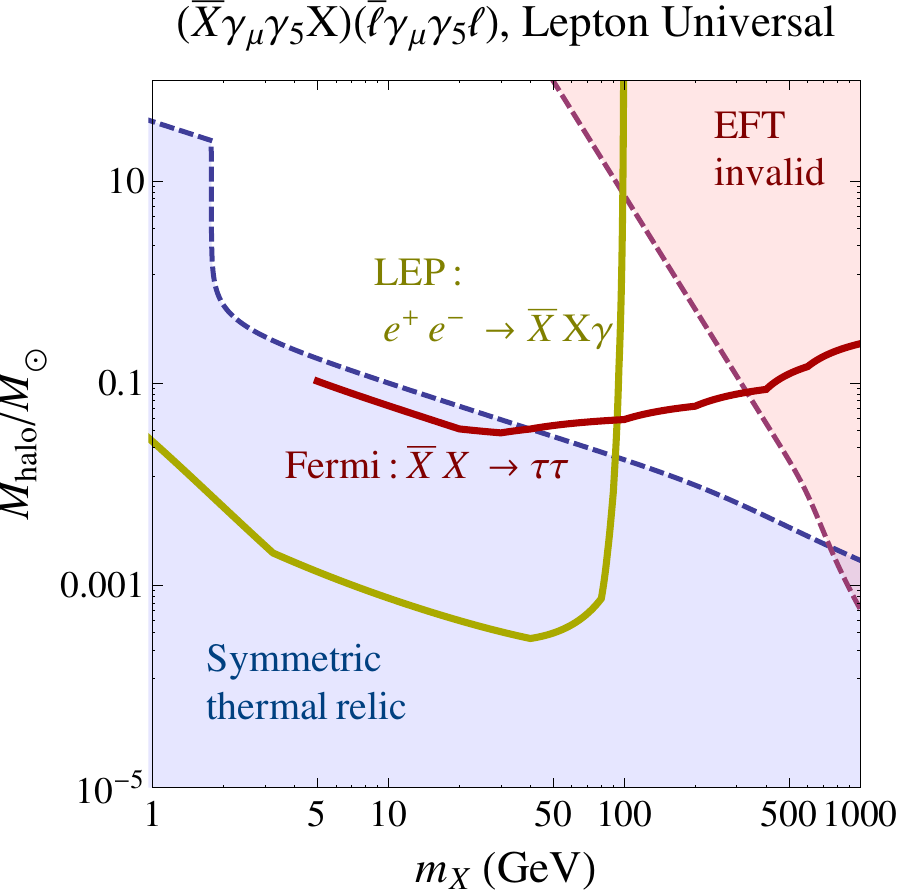}
\caption{LEP limits (yellow) apply to the electron-type coupling and are based on monophoton + MET searches~\cite{Fox:2011fx}. The Fermi-LAT limit (dark red) obtained from the joint likelihood analysis of 10 dwarf galaxies for DM annihilating to $\overline{\tau}\tau$~\cite{Ackermann:2011wa}.  For comparison we show the thermal symmetric WIMP (dashed blue line) for $\overline{X}X \rightarrow \ell^{+} \ell^{-}$ comprising the total annihilation cross section. Thermal symmetric models lie within the blue region, whereas thermal asymmetric models lie above. We have highlighted in pink the region, $\Lambda > m_{X}/2\pi$, where the assumption of EFT is no longer valid.}
\label{ellDM}
\end{center}
\end{figure*}

\label{types}

The types of DM-lepton interactions we will study are vector, scalar and axial vector operators:
\bea \mathcal{O}_{V}&=& \frac{1}{\Lambda^{2}}\left(\overline{X} \gamma_{\mu}X \right) \left(\overline{\ell}\gamma^{\mu} \ell\right), \nonumber \\
 \mathcal{O}_{S} &=& \frac{1}{\Lambda^{2}}\left(\overline{X}X \right) \left(\overline{\ell} \ell\right).\\
 \mathcal{O}_{A} &=& \frac{1}{\Lambda^{2}}\left(\overline{X} \gamma_{\mu}\gamma^{5}X \right) \left(\overline{\ell}\gamma^{\mu}\gamma^{5} \ell\right) \nonumber,
\eea
where $\Lambda$ is the interaction mass scale.  We will turn on each operator, one a time to derive constraints. With this set of assumption the only parameters are the DM mass $m_{X}$ and the interaction scale $\Lambda$.

With the exception of the LEP missing energy limits, our constraints will come from DM annihilation searches. Thus the precise form of the annihilation cross section will determine the sensitivity to $\Lambda$. With non-relativistic DM annihilation our cross sections are:
\bea
\langle \sigma_{ann} v \rangle_{V} &=& \frac{m_{X}^{2}}{ 2\pi \Lambda^{4}} \sum_{f} \left(1- \frac{m_{f}^{2}}{m_{X}^{2}}\right)^{1/2} \left(2+ \frac{m_{f}^2}{m_{X}^{2}}\right) \nonumber \\
 \langle \sigma_{ann} v \rangle_{S} &=& \frac{m_{X}^{2}v^{2}}{ 8\pi \Lambda^{4}} \sum_{f} \left(1- \frac{m_{f}^{2}}{m_{X}^{2}}\right)^{1/2} \\
 \langle \sigma_{ann} v \rangle_{A} &=& \frac{1}{ 2\pi \Lambda^{4}} \sum_{f} m_{f}^{2} \left(1- \frac{m_{f}^{2}}{m_{X}^{2}}\right)^{1/2}, \nonumber
\eea
where $m_{X}$ and $m_{f}$ are the DM and lepton masses.  With these three operator choices, we have $s$-wave, $p$-wave suppressed and helicity suppressed annihilation respectively.

The other relevant piece operator-dependent physics is elastic scattering, which sets the temperature of kinetic decoupling.  Given that the DM is non-relativistic by the era of kinetic decoupling, the concomitant momentum transfer in this era is quite small.  Thus typically one only need worry about the forward scattering cross section. This approximation is widespread in the literature~\cite{Bringmann:2009vf}, however it has been pointed out that the use of this approximation leads to a small overestimate of the actual momentum relaxation rate, and thus slightly underestimates the temperature of kinetic decoupling~\cite{Gondolo:2012vh}. In what follows (see Sec.~\ref{KD}), we shall see that kinetic decoupling is determined by the quantity
\be \overline{\left(\sigma_{fX} t \right)} \equiv \int_{-4p^{2}}^{0} dt \left(-t\right) \frac{d\sigma_{fX}}{dt}
= \alpha_{\mathcal{O}} \frac{p^{4}}{\Lambda^{4}},
\label{cx}
\ee
where $\sigma_{fX}$ is the elastic scattering cross section of DM $X$ on a SM particle $f$, and $\alpha_{\mathcal{O}}$ is an operator dependent numerical coefficient given by
\be \alpha_{V} = \frac{2}{3\pi},~~\alpha_{S} = \frac{1}{48\pi}, ~~\alpha_{A} = \frac{10}{3\pi}.
\ee

Lastly, let us comment on the possibility of having very strong DM-neutrino interactions without DM-charged lepton interactions of the same strength. One may be worried that this cannot be accomplished since neutrinos and charged lepton come in the same doublet of $SU(2)_{L}$. However, this concern can be ameliorated at the level of effective field theory by considering the dimension-8 coupling $\mathcal{O}_{LH} \mathcal{O}_{X}$, where $\mathcal{O}_{\psi}$ refers to some Lorentz- and gauge- invariant bilinear of $\psi$. Thus upon electroweak symmetry breaking, the Higgs acquires a vacuum expectation value, and neutrino-DM interactions are induced without same strength charged lepton-DM interactions (see~\cite{Berezhiani:2001rs} for the analogous argument involving BSM quark-neutrino interactions).  

\section{Constraining $\ell$DM interactions}
\label{exp}

A number of experiments are sensitive to lepton-dark matter ($\ell$DM) interactions.  Those that we will be especially interested in are:
\begin{itemize}
\item {\bf  LEP}: Missing energy searches with a single photon at LEP provide stringent limits on BSM physics~\cite{Berezhiani:2001rs,Fox:2011fx}.   These mono-photon searches provide constraints on DM masses that are kinematically accessible, $m_{X} \lesssim 100$. These limits are in fact sufficiently strong to exclude thermal DM annihilating to electrons for masses below 20 GeV or so~\cite{Fox:2011fx}.  

\item{\bf Fermi-LAT}: Direct searches for DM annihilating to $\mu^{+}\mu^{-}$ and $\tau^{+}\tau^{-}$ are encroaching on light WIMP territory as well, having already excluded thermally sized cross sections below roughly 40 and 30 GeV respectively. Here we use null results from the joint likelihood analysis of 10 dark matter-dominated dwarf galaxies with 24 months of data~\cite{GeringerSameth:2011iw,Ackermann:2011wa}. 

\item{\bf Direct detection}: Until very recently most direct detection searches vetoed events from electronic scattering. Now however, using the methods outlined in~\cite{Essig:2011nj}, XENON10 has placed the first limits on such scattering for DM masses in the MeV to GeV mass range~\cite{Essig:2012yx}. Though the limits are quite mild at present (being based on only 15 kg-days of exposure), such an analysis represents an important proof-of-principle for future studies of DM-electron interactions.

\item {\bf IceCube}: Recently, very large neutrino telescopes have begun deriving stringent limits on neutrino-DM interactions. IceCube~\cite{Abbasi:2011eq,Abbasi:2012ws} now provides direct constraints on the DM annihilation cross section into neutrinos. Though their current sensitivity remains orders of magnitude away from the thermal relic scale, substantial improvements are likely to follow with the 79 string upgrade to IceCube, and it will likely be competitive with the unpublished results of Super-K.

Here we will use the results of the 22-string analysis~\cite{Abbasi:2011eq}, rather than the somewhat weaker limits arising from the 40-string analysis~\cite{Abbasi:2012ws}.  Note that although there is substantial uncertainty in the Galactic Center limit due to the choice of galactic halo model, the IceCube collaboration removes the Galactic Center from the analysis such that the resulting bound is due to DM in the outer halo region where model uncertainties are much smaller. 

\item{\bf Super-K}: The Super-Kamiokande collaboration has an unpublished galactic halo neutrino line search~\cite{SKGalactic}. We include these results for reference here, though they may of course change significantly once the results are finalized.

\item{\bf CMB}: If DM remains in thermal equilibrium with neutrinos at temperature below $\mathcal{O}({\rm MeV})$, it can be constrained by the recent Planck data through its affect on the effective number of neutrinos, $N_{{\rm eff}}$~\cite{Boehm:2013jpa}. We make use of the constraint obtained from Planck data that excludes (at 95$\%$ CL), Dirac DM with a mass $<8.7$ MeV in equilibrium with neutrinos.

\end{itemize}
%

\section{Kinetic Decoupling and Dark Matter Protohalos}
\label{KD}

Two distinct physical processes set the size of the smallest DM structures: free streaming~\cite{Green:2005fa} and acoustic damping~\cite{Loeb:2005pm,Bertschinger:2006nq}. After the comoving number density of DM has ``frozen out'' and chemical equilibrium has been lost, the elastic scattering of DM on the thermal bath continues the efficient exchange of momentum such that kinetic equilibrium persists until late times. This process effectively damps the growth of perturbations that would otherwise grow to form the first DM subhalos. Once DM decouples, it can stream freely from overdense regions into underdense regions and efficiently erase structure on small scales. In general, it is the largest of these two physically independent processes that sets the scale of the cut-off in the power spectrum. We therefore define this cutoff as
\be M_{halo} \equiv \max \left(M_{FS},M_{KD}\right)
\ee
When DM kinetically decouples, it can stream freely until matter-radiation equality when structure formation begins in full force. This process significantly damps fluctuations below the scale of free-streaming $k_{fs}$ such that the smallest protohalo allowed by free-streaming is~\cite{Bringmann:2009vf}:
\bea 
M_{FS} &\approx& 2.9 \times 10^{-6}M_{\odot}\\
&\times& \left(\frac{1+ \log \left(g_{eff}^{1/4}T_{KD}/50~{\rm MeV}\right)/19.4}{\left(m_{X}/100\right)^{1/2} g_{eff}^{1/4} \left(T_{KD}/50~{\rm MeV}\right)^{1/2}}\right)^{3} \nonumber
\label{MFS}
\eea

Moreover the damping scale set by acoustic oscillations is given by the DM mass enclosed in the horizon at this epoch,
\bea  &M_{KD}&(T_{KD}) = \frac{4\pi}{3}\frac{\rho_{X}(T_{KD})}{H(T_{KD})^{3}} \\
 & \approx & 9\times 10^{-7}M_{\odot} \left[\frac{ h(T_{KD})}{g(T_{KD})^{3/2}T_{KD}^{3}}\right]_{T_{KD}=50~{\rm MeV}}. \nonumber
\label{MKD}
\eea
Thus we see that as the DM decreases, the mass scale set by free-streaming becomes increasingly important. In general however, it is useful to notice that the acoustic oscillation scale (Eq.~\ref{MKD}) falls off more rapidly with the decoupling temperature than free-streaming (Eq.~\ref{MFS}). Thus at sufficiently low-temperatures, the cutoff in the power spectrum will always be set by acoustic oscillations.

Importantly, after the QCD phase transition, the number density of pions and hadrons becomes so small that kinetic equilibrium is only maintained by lepton-DM interactions. (Indeed it has been recently shown that the inclusion of DM-pion scattering modifies $T_{KD}$ by less than one percent in most cases~\cite{Gondolo:2012vh}.)  One can immediately see that as long as acoustic oscillations dominate the damping, one expects the size of the protohalos are constrained to be $\lesssim 10^{-7}M_{\odot}$ for DM models with quark-only couplings.

The temperature of kinetic decoupling can be estimated roughly from simple analytic arguments. These assumptions are accurate when the bath scattering partners are relativistic and when the degrees of freedom are not changing rapidly~\cite{Bringmann:2006mu}. Similar treatments to the methods we follow can be found in~\cite{Schmid:1998mx,Bringmann:2006mu,Bringmann:2009vf}. 

Let us begin with a simple sketch of the physics relevant for the kinetic decoupling of DM.  At the temperatures relevant for kinetic decoupling, DM is non-relativistic. Thus the average change in DM momentum from a single scattering event with a relativistic lepton is $\Delta p \approx T$ is small compared to the DM equipartition momentum, $\sqrt{m_{X}T}$.  For DM to remain in kinetic equilibrium however the fractional change in momentum must be $\mathcal{O}(1)$. Since bath scatterings induce a random walk in momentum space, we require $N \approx m_{X}/T$ collisions to maintain equilibrium and ensure large fractional change in momentum. We can therefore simply estimate the temperature when DM loses kinetic equilibrium with a bath species $f$ by solving 
\be \gamma_{p}(T_{KD}) \simeq H(T_{KD}),
\label{hubble}
\ee
where the momentum relaxation rate can be roughly approximated by, $\gamma_{p}(T_{KD}) \simeq (T/m_{X}) n_{f} \sigma_{Xf}$, with $n_{f}$ the number density of the species $f$, and $\sigma_{Xf}$ the $t=0$ scattering cross section of DM on $f$.  The above arguments make clear the physical effects that enter into a determination of the momentum relaxation rate, and provide a intuitive picture of its parametric dependence. 

However, an improved estimate of $\gamma_{p}$ can be obtained from the Fokker-Planck equation. In what follows we will use the results of~\cite{Gondolo:2012vh}, which follow this method and include the effects of Pauli blocking and contributions from the non-forward parts of the cross section.  Under the assumption that the bath particles are relativistic the momentum relaxation rate is well-approximated by

\begin{widetext}
\be 
\gamma_{p}(T) = \frac{g_{f}}{6m_{X}T} \int_{0}^{\infty} \frac{d^{3}p}{\left(2 \pi \right)^{3}}  f(p/T) (1-f(p/T)) \int_{-4p^{2}}^{0} dt\left(-t\right) \frac{d\sigma_{X f}}{dt},
\label{gam}
\ee
\end{widetext}
where $g_{\ell}$ are the internal degrees of freedom of the bath particle, $f(p/T)$ is the phase space occupancy function, and $p$ is the center of mass momentum. 

Finally, we can solve for the temperature of kinetic decoupling by combining Eqs.(\ref{cx}),(\ref{gam}), and (\ref{hubble}) to arrive at
\be	
T_{KD} \simeq   \frac{0.69 g_{eff}^{1/8}}{g_{f}^{1/4}} \left(\frac{\Lambda^{4} m_{X}}{M_{Pl}\alpha_{\mathcal{O}} }\right)^{1/4}\left(\frac{T}{T_{\nu}}\right)^{1/2},
\label{T2}
\ee
where $g_{f}$ are the internal degrees of freedom of the bath particle $f$, and $g_{eff}$ are temperature-dependent energy degrees of freedom.  Note that we have been careful to include the possibility that the kinetic decoupling of DM may occur after neutrinos have decoupled, for which $T\neq T_{\nu}$. In what follows we will use this expression for the kinetic decoupling temperature to translate constraints on the interaction scale $\Lambda$ into constraints on protohalo mass.

\section{Results}
\label{lepDM}
\subsection{Charged Lepton Constraints}
\begin{figure}[t] 
\begin{center}
 \includegraphics[width=\columnwidth]{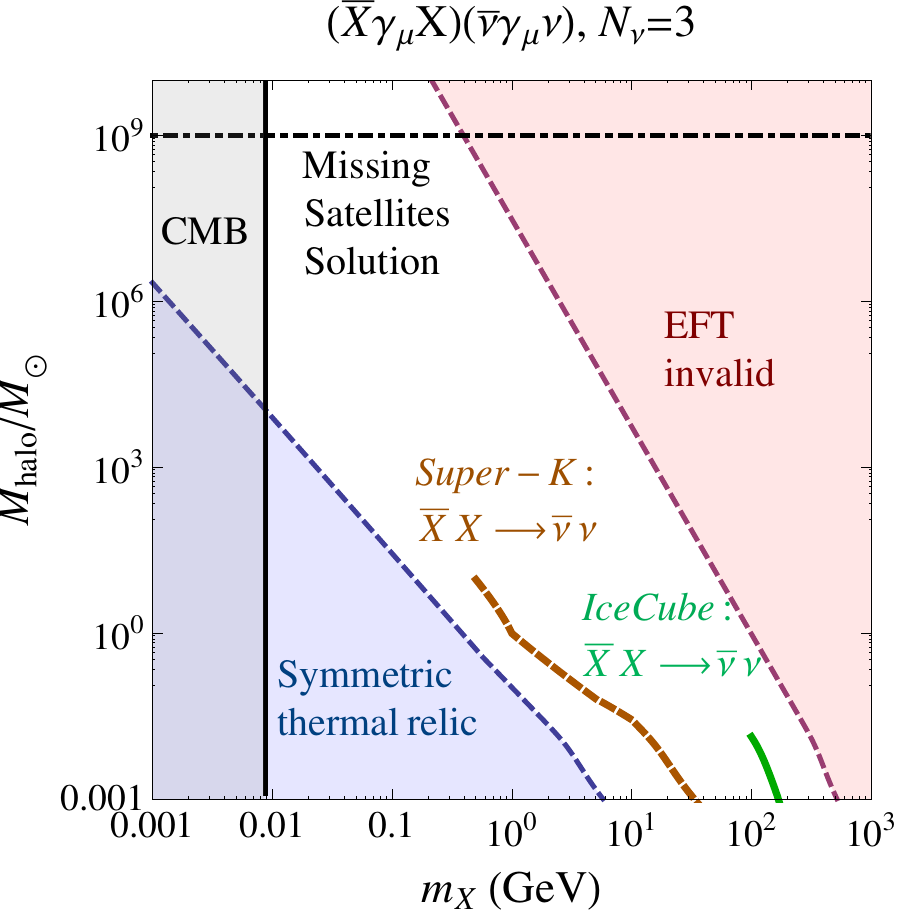}
\caption{Constraints on the protohalo mass for neutrinophilic DM.   Experimental upper bounds come from IceCube's galactic halo analysis~\cite{Abbasi:2011eq} which restricts DM annihilation into neutrinos (green, solid) and Super-K which has a similar but unpublished preliminary search~\cite{SKGalactic} that extends to lower masses (dashed dark orange). The CMB constraint comes the Planck limit on $N_{\rm eff}$ as analyzed in~\cite{Boehm:2013jpa}.}
\label{nuDM}
\end{center}
\end{figure}

Let us finally come to the main results of this paper, summarized in Figs. \ref{ellDM} and \ref{nuDM}. Figure~\ref{ellDM} shows how the relevant constraints on DM-charged lepton interactions translate into constraints on the DM protohalo mass. Here we have chosen universal couplings to all leptons.  For reference, in each plot, we have also included the thermal relic density constraint $\langle \sigma v \rangle_{sym} = 6 \times10^{-26}~{\rm cm}^{3}{\rm s}^{-1}$ for Dirac DM, assuming only one given operator is turned on at a time. We must highlight a few key points about this region. First, consider a relic species in the presence of zero particle-antiparticle asymmetry.  In this case, the correct thermal relic abundance is obtained {\it entirely} by the specified operator for points that line on the blue dashed line. Of course, in general, one expects more than one annihilation mode to contribute, in which case any given mode must not exceed the total required annihilation cross section, $\langle \sigma v \rangle_{sym}$. Thus symmetric thermal relic models, lie above the blue dashed curve.  However, now consider a model in which a nonzero asymmetry exists. Then, as long we do not revoke the thermal hypothesis, the total annihilation cross section must be greater than the $\langle \sigma v \rangle_{sym}$~\cite{Graesser:2011wi}. Thus for asymmetric models with just one operator turned on, the correct abundance requires DM to lie above the blue curve. 

We now turn to the constraints on the vector operator, $\mathcal{O}_{V}$, displayed in the top panel of Figure~\ref{ellDM}. There we see that the limits are quite strong, especially at high DM masses. This can be traced to the fact that $s-$wave annihilation is strongly bounded by the Fermi data.  At the lower mass end, where the LEP center of mass energy is large compared to the energy required to produce a DM pair, the constraints are also strong.  However it is important to notice that the LEP limit on the protohalo mass becomes increasingly stringent as $m_{X}$ increases until around masses $\mathcal{O}(100)$ GeV, despite the fact the limit on $\Lambda$  degrades as $m_{X}$ approaches 100 GeV. Inspecting Eqs.~(\ref{MKD}) and (\ref{T2}) however, one sees that this is accounted for by the fact that the mass enclosed in the horizon at kinetic decoupling scales as $M_{KD} \propto m_{X}^{-3/4}$. Below masses of around 5 GeV, the mass scales set by free-streaming and acoustic oscillations cross over leading to the noticeable kink in the LEP curve. However, the kink in the EFT curve around a  few hundred GeV is due to the change in the relativistic degrees of freedom when $e^{\pm}$ pairs become Boltzmann suppressed.  A more accurate treatment including the effects of non-relativistic bath particles should smoothen this effect.

The scalar operator, $\mathcal{O}_{S}$ (displayed in bottom left panel of Fig.~\ref{ellDM}) is quite different.  In this case, there are two qualitative differences from $\mathcal{O}_{V}$: (1) annihilation is $p$-wave suppressed (substantially weakening the Fermi limits) (2) the DM-bath elastic scattering is weaker than in the vector case.  The LEP limits are not especially sensitive to the Lorentz structure of the operator, so the concomitant limits on the cutoff $\Lambda$ are nearly the same as they are for $\mathcal{O}_{V}$ ~\cite{Fox:2011fx}. However, since the elastic scattering of DM on leptons is weaker, the corresponding constraints on the protohalo mass are {\it stronger}~ since thermal decoupling occurs earlier.


Finally, the axial vector case, $\mathcal{O}_{A}$, is displayed in the bottom right panel of Fig.~\ref{ellDM}. Here the DM has helicity suppressed annihilation and elastic scattering qualitatively much like the vector case. The helicity suppression implies that the annihilation into tau leptons dominates over nearly the entire mass range (this is the origin of the bump near 2 GeV where tau annihilation transitions to muon annihilation). Since LEP is insensitive to the Lorentz structure and the elastic scattering proceeds just as for $\mathcal{O}_{V}$, the resultant LEP constraints are nearly the same.  The observed rise in the limit on $M_{halo}$ from Fermi data is simply due to the fact the Fermi's sensitivity to DM annihilation weakens with increasing DM mass. This rise does not occur for $\mathcal{O}_{V}$ since two powers of $m_{X}$ in the annihilation cross section compensate this effect.

In total, the constraints on the protohalo mass implied by the combination of LEP and Fermi data are relatively strong and do not allow any potential solutions to the missing satellite problem. In what follows, we will examine a way around this conclusion.

\subsection{Neutrinophillic DM}
In view of the above stringent constraints, we ask: how they can be weakened? The simplest possibility is to permit only neutrino-DM couplings. Neutrinophilic DM ($\nu$DM) of this type has been studied by many authors, e.g.~\cite{Boehm:2000gq,Mangano:2006mp,Boehm:2006mi,Hooper:2007tu,Boehm:2013jpa}.

Let us examine the constraints on the operator $\left(\overline{X} \gamma_{\mu}X \right)\left(\overline{\nu} \gamma_{\mu}\nu \right)$. In the top panel of Fig.~\ref{nuDM} we display the resultant constraints from Planck, IceCube, and Super-Kamiokande. Both of the neutrino line searches are based on their galactic halo analyses, and require no other DM interactions. Of course, Solar searches for DM annihilation can be relevant when DM also has interactions with nuclei sufficient to induce gravitational capture by the Sun. We do not include these constraints here.

The constraints displayed in Fig.~\ref{nuDM} reveal that the assumption of EFT is the limiting factor at low masses. Recall that the missing satellite problem could indeed be an indication that the DM power spectrum has a cutoff at large masses. Then, $\nu$DM with a heavy mediator could account for such a large cutoff only if DM is sub-GeV in mass. This range of masses and interaction strengths, corresponds to very large annihilation rates. A thermal relic abundance with a larger than WIMP value can easily be accommodated in models with a nonzero particle/antiparticle asymmetry~\cite{Graesser:2011wi}. 

Of course, effective $\nu$DM cannot have arbitrarily large DM protohalos. Although we do not display the Lyman-$\alpha$ constraints (see for example~\cite{Boyarsky:2008xj}) on the protohalo mass, they imply that this mass cannot be larger than $ 5\times 10^{10}~M_{\odot}$.

UV complete models of $\nu$DM may also offer a solution to the missing satellites problem even at large DM masses~\cite{Aarssen:2012fx}. The minimal model introduced in~\cite{Aarssen:2012fx} explicitly breaks gauge invariance and hence suffers from very strong constraints from kaon and W decays~\cite{Laha:2013xua}. Such constraints are however strongly model-dependent, not applying for example in models where the mediator couples dominantly to sterile neutrinos. We hope to return to this problem in future work.

\section{Discussion and Conclusions}
\label{conc}
In this paper we have attempted to systematically study how constraints on lepton-DM interactions map into protohalo constraints. We have shown that surprisingly strong limits on the DM protohalo mass are obtained in the regime where EFT is valid and when charged lepton couplings exist.

If the missing satellite problem is indicative of non-standard DM physics, it would likely indicate that DM is neutrinophilic. Generically in the heavy mediator limit we have considered here, large protohalo masses requires very light DM masses, though this is unlikely to be true when one revokes the effective field theory assumption, and allows for the possibility of light mediators. The topic of light mediators connecting DM and neutrinos deserves further work in light of this work (see~\cite{Aarssen:2012fx} for an example). 

Future constraints on the coupling of DM to both charged leptons and especially neutrinos will improve on the limits derived here.  Potential future constraints on charged lepton-DM interactions will come from direct detection searches for electron-DM scattering, upcoming Fermi-LAT releases, a re-analysis of DM limits from solar annihilation when DM couples only to electrons, and future electron-positron collider limits~\cite{Chae:2012bq}.  

The strength of $\nu$DM interactions will be further limited by low-threshold neutrino line searches. Thus existing Super-K data and future DeepCore/PINGU~\cite{Koskinen:2011zz} results from IceCube will be critical in determining whether or not DM is neutrinophilic.

\acknowledgements
\vspace{-.5cm}
We are especially grateful to Alex Friedland, Michael Graesser, Christopher McCabe, Kalliopi Petraki, Carsten Rott, and Luca Vecchi for enlightening and lively conversations. We would also like to thank the organizers of the Kavli Institute for Theoretical Physics program, {\it Hunting for Dark Matter: Building a cross-disciplinary, multi-pronged approach}, where a portion of this work was performed. 

This research was supported by the LANL LDRD program and the National Science Foundation under Grant No. NSF PHY11-25915.
 
\bibliography{nuDM}
\end{document}